\documentclass[prl,reprint,superscriptaddress,longbibliography,floatfix,twocolumn,nofootinbib,notitlepage]{revtex4-1}
\usepackage[utf8]{inputenc}
\usepackage{graphicx}
\usepackage[normalem]{ulem}
\usepackage{amsmath,amssymb,amsthm,bm,mathtools,amsfonts,mathrsfs,bm,bbm,dsfont}
\usepackage{physics,xcolor}

\newcommand{\be}{\begin{equation}}
\newcommand{\ee}{\end{equation}}

\newcommand{\prjct}[1]{\mathinner{|{#1}\rangle}\!\!\mathinner{\langle{#1}|}}

\newcommand{\id}{\mathds{1}}
\newcommand{\ii}{\mathrm{i}}

\newcommand{\cE}{\mathcal{E}}

\newcommand{\cH}{\mathcal{H}}

\renewcommand{\leq}{\leqslant}
\renewcommand{\geq}{\geqslant}

\newtheorem{claim}{Claim}

\begin{document}

\title{Simulability of high-dimensional quantum measurements}

\author{Marie Ioannou}
\thanks{These authors contributed equally to this work}
\affiliation{Department of Applied Physics University of Geneva, 1211 Geneva, Switzerland}
\author{Pavel Sekatski}
\thanks{These authors contributed equally to this work}
\affiliation{Department of Applied Physics University of Geneva, 1211 Geneva, Switzerland}
\author{S\'ebastien Designolle}
\affiliation{Department of Applied Physics University of Geneva, 1211 Geneva, Switzerland}
\author{Benjamin D.M. Jones}
\affiliation{Department of Applied Physics University of Geneva, 1211 Geneva, Switzerland}
\affiliation{H. H. Wills Physics Laboratory, University of Bristol, Bristol, BS8 1TL, UK}\affiliation{Quantum Engineering Centre for Doctoral Training,  University of Bristol, Bristol, BS8 1FD
UK}\author{Roope Uola}
\affiliation{Department of Applied Physics University of Geneva, 1211 Geneva, Switzerland}

\author{Nicolas Brunner}
\affiliation{Department of Applied Physics University of Geneva, 1211 Geneva, Switzerland}

\begin{abstract}
    We investigate the compression of quantum information with respect to a given set $\mathcal{M}$ of high-dimensional measurements. This leads to a notion of simulability, where we demand that the statistics obtained from $\mathcal{M}$ and an arbitrary quantum state $\rho$ are recovered exactly by first compressing $\rho$ into a lower dimensional space, followed by some quantum measurements. A full quantum compression is possible, i.e., leaving only classical information, if and only if the set $\mathcal{M}$ is jointly measurable. Our notion of simulability can thus be seen as a quantification of measurement incompatibility in terms of dimension. After defining these concepts, we provide an illustrative examples involving mutually unbiased basis, and develop a method based on semi-definite programming for constructing simulation models. In turn we analytically construct optimal simulation models for all projective measurements subjected to white noise or losses. Finally, we discuss how our approach connects with other concepts introduced in the context of quantum channels and quantum correlations.
\end{abstract}

\maketitle

\paragraph{Introduction.---} Quantum measurements play a fundamental role in quantum theory and its applications, notably in quantum information processing and metrology. Indeed measurements represent the bridge between a quantum system and an external observer, hence essentially any quantum experiment relies on a quantum measurement process. More recently, the role of quantum measurements as a resource was clarified in the context of quantum information processing tasks, see \cite{JMinvitation,JMreview} for a review. At the formal level, the concept of joint measurability \cite{busch16} provides a framework for characterisation and quantification of the incompatibility of quantum measurements, which can be connected to their usefulness in, e.g., state-discrimination problems \cite{carmeli19a,skrzypczyk19,oszmaniec19,uola19b,uola19c,Ducuara2020} and quantum steering \cite{quintino14,uola14,uola15,kiukas17}.

A natural question is whether quantum measurement incompatibility can also be quantified in terms of dimension. Intuitively, a set of quantum measurement defined on a high-dimensional Hilbert space may feature a stronger form of incompatibility than what is possible for lower dimensions. 

In this work we address this question and propose a notion of dimensionality for measurement incompatibility. This notion can be understood, and naturally motivated, in a scenario involving the compression of quantum information. Loosely speaking, we ask whether the statistics of a set of $d$-dimensional positive operator-valued measures (POVMs) $\mathcal{M}$ (considering any possible quantum state $\rho$) can be exactly recovered from first projecting $\rho$ onto a {$n$-dimensional} space (with $1 \leq n<d$) and then performing POVMs in this lower-dimensional space. If such a compression is possible, we say the set $\mathcal{M}$ is $n$-simulable. Note that the case $n=1$ exactly corresponds to the notion of joint measurability; indeed, in this case, the full quantum information can be compressed to classical information, see also \cite{heinosaari2017,guerini2019distributed}. However, as we will see below, there exist sets of POVMs $\mathcal{M}$ that are incompatible, but yet simulable with $n$-dimensional measurements with $1<n<d$. The notion of $n$-simulability can thus be seen as an extension of the concept of joint measurability, providing a quantification of the incompatibility of quantum measurements in terms of dimension.

After introducing these ideas more formally, we provide illustrative examples based on sets of mutually unbiased measurements. Then, we present a method based on semi-definite programming (SDP) to show that a set of measurements is $n$-simulable. In turn we consider the continuous sets of all projective measurements subjected to noise or losses, and construct optimal $n$-simulation models. Finally, we establish a link to partially entanglement-breaking channels \cite{chruscinski2006partially}, and discuss connections to other concepts as well as open questions.

\emph{Scenario and definition.---} Consider the following task: a sender (Alice) is located on the moon and wants to transmit an (arbitrary) $d$-dimensional quantum state $\rho$ to a receiver (Bob) located on earth. Upon receiving $\rho$, Bob will perform a set of measurements $\mathcal{M} = \{M_{a|x}\}_{a,x}$, where $\{ M_{a|x}\}_a$ denotes a POVM, i.e., $M_{a|x}\geq0$ and $\sum_a M_{a|x} = 1$ $\forall a,x$. This leads to the following statistics termed the target data: $p(a|x,\rho) = \Tr(M_{a|x}\rho)$.

\begin{figure}[h]
    \centering
    \includegraphics[width=0.95\columnwidth]{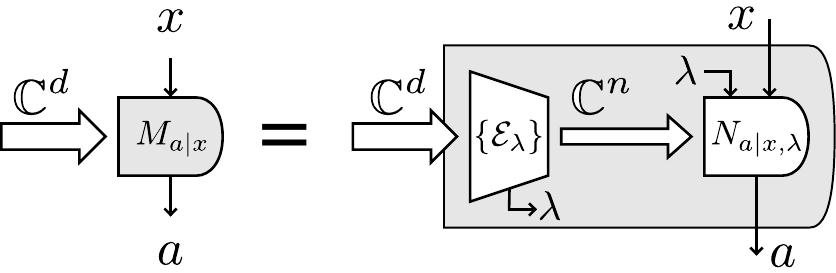}
    \caption{A measurement assemblage $\{M_{a|x}\}$ in dimension $d$ is said to be $n$-simulable if it can be replicated by first compressing the measured system down to dimension $n<d$ with a ``parent'' instrument $\{\cE_\lambda\}$ independent of the setting $x$, and then performing some measurements $\{N_{a|x,\lambda}\}$ on the $n$-dimensional system.}
    \label{fig:1}
\end{figure}

So far, we assumed that the channel between Alice and Bob is ideal, i.e., a $d$-dimensional identity channel. In the following, however, we will consider the dimensionality of the quantum channel as a resource, which we will aim to minimise. Note that a classical channel (of arbitrary capacity) is always available for free.

We now ask if the target data could be reproduced by using a lower-dimensional quantum channel. That is, we look for a quantum instrument $\{\cE_\lambda\}_\lambda$, where each $\cE_\lambda: B(\mathbb{C}^d) \to B(\mathbb{C}^n)$ is a completely positive (CP) map, $\lambda$ a classical outcome, and the map $\sum_\lambda \cE_\lambda$ is trace-preserving. Alice would then send the compressed $n$-dimensional state $\cE_\lambda (\rho)$ to Bob, together with the classical outcome $\lambda$. Bob finally chooses from a set of measurements $\mathcal{N}=\{N_{a|x,\lambda}\}_{a,x,\lambda}$. The protocol is successful if we recover the target data, i.e., if
\be \label{match}
  \sum_{\lambda}\Tr[N_{a|x,\lambda} \cE_\lambda (\rho)]  = \Tr(M_{a|x}\rho)   \quad \forall \rho \, ,
\ee
or equivalently if $M_{a|x} = \sum_{\lambda} \cE_\lambda^*( N_{a|x,\lambda}) $, where $\cE_\lambda^*$ is the adjoint map to $\cE_\lambda$ (for continuous instruments one replaces the sum with an integral, see examples below). In this case, we say that $\mathcal{M}$ is $n$-simulable, as illustrated in Fig.~\ref{fig:1}. Clearly, in the case $n=1$, the instrument corresponds to a POVM and the measurements afterwards form classical post-processings of the classical output. This coincides with the concept of joint measurability: $M_{a|x}=\sum_\lambda G_\lambda p(a|x,\lambda)$, where $\{G_\lambda\}$ represents the joint (or parent) POVM \cite{JMinvitation,JMreview}.

\paragraph{Illustrative example.---} To illustrate the concept, we present an example involving mutually unbiased bases (MUBs). 
Recall that two bases $\{\ket{\varphi_i^1}\}_i$ and $\{\ket{\varphi_j^2}\}_j$ are termed mutually unbiased if $|\braket{\varphi_i^1}{\varphi_j^2}|^2 = 1/d$ for all $i$ and $j$. There are at most $d+1$ MUBs in dimension $d$, and a construction for the complete set of $d+1$ MUBs is only known when $d$ is a power of a prime number~\cite{WF89}. 

Here we consider a set $\mathcal{M}$ consisting of $m$ measurements in MUBs subject to white noise.
That is, projection-valued measures (PVMs) composed of projectors $P_{a|x} = \ketbra{\varphi_a^x}{\varphi_a^x}$ preceded by a white noise channel. The resulting POVMs read
\begin{equation}\label{eqn:Meta}
    M_{a|x}^\eta=\eta P_{a|x}+(1-\eta)
    \frac{\id}{d},
\end{equation}
where $x \in \{1,\ldots,m\}$ and $a \in \{1,\ldots,d\}$. Note that we use the standard construction of MUBs \cite{WF89} with $\{\ket{\varphi_i^1}\}_i$ being the computational basis.

We now show that, depending on the amount of noise, $1-\eta$, the set $\mathcal{M}$ becomes $n$-simulable. Let us first construct an appropriate map, implementing the compression from dimension $d$ to $n$. For a given basis $\{\ket{\psi_i}\}_i$ we consider the set of $d \choose n$ projectors $\Pi_\lambda$ onto an $n$-dimensional subspace of $\mathbb{C}^d$ spanned by the vectors of the basis.

This defines an instrument $\{\cE_\lambda\}$, with each $\cE_\lambda$ given by a Kraus operator $K_\lambda = \sqrt{\binom{d}{n}^{-1} \frac{d}{n}} \, \Pi_\lambda$, compressing from dimension $d$ to $n$. Here $\binom{d}{n}$ is simply the number of projectors $\Pi_\lambda$ with $n$ ones on the diagonal, and the term $\sqrt{\frac{d}{n}}$ is due to normalisation.
Next, for any POVM $P_a$ acting on $\mathbb{C}^d$, define its restriction to the subspace labelled by $\lambda$ as
\be\label{eq: compressed gen}
N_{a|\lambda} = \Pi_\lambda P_a \Pi_\lambda,
\ee
which is a POVM on $\mathbb{C}^n$ since $N_{a|\lambda} \geq 0$ and $\sum_a N_{a|\lambda} = \Pi_\lambda \id_d \Pi_\lambda = \id_n$. We can now compute 
\be\begin{split}
    \sum_\lambda \cE^{*}_\lambda(N_{a|\lambda})=
     \frac{n-1}{d-1} P_a + \left(1-\frac{n-1}{d-1}\right) \mathcal{T}_{\{\ket{\psi_i}\}} [P_a]
\end{split}
\ee
where $\mathcal{T}_{\{\ket{\psi_i}\}} [P_a] =  \sum_{i} \ketbra{\psi_i}{\psi_i} P_a \ketbra{\psi_i}{\psi_i}$ is the twirling map in the basis $\{\ket{\psi_i}\}$ used to define the instrument. In particular, if the eigenbasis of $P_a$ and $\{\ket{\psi_i}\}$ are mutually unbiased, one obtains $\mathcal{T}_{\{\ket{\psi_i}\}} [P_a] =\frac{\id}{d}$. While if they coincide, one trivially gets $\mathcal{T}_{\{\ket{\psi_i}\}} [P_a] = P_a$.

If the set $\mathcal{M}$ is composed of $m$ MUBs, we construct an instrument that chooses one of the bases $y=1,\dots,m$ randomly and performs $\{\cE_{\lambda|y}\}$ as defined above. Then, for the setting $x$ Bob does the measurement $N_{a|\lambda,x,y} $ as defined in Eq.~\eqref{eq: compressed gen}. This construction results in
\be\begin{split}
&\sum_{\lambda,y} \frac{1}{m}\cE_{\lambda|y}^*(N_{a|\lambda,x,y}) = \frac{1}{m} P_{a|x} \\ &+ \frac{m-1}{m}\left[\frac{n-1}{d-1} P_{a|x} + \left(1-\frac{n-1}{d-1}\right)  \frac{\id}{d} \right],
\end{split}
\ee
which equals $M^\eta_{a|x}$ for $1-\eta =\frac{m-1}{m}\left(1-  \frac{n-1}{d-1}\right)$, and implies the following observation.

\begin{claim} The set $\mathcal{M}$ in Eq.~\eqref{eqn:Meta} of $m$ noisy projective measurements in MUBs on $\mathds{C}^d$ is $n$-simulable if
\be
\label{eqn:claim1}
\eta \leq 1- \frac{m-1}{m}\cdot\frac{d-n}{d-1}.
\ee
\end{claim}

Let us first discuss the case of a pair of MUBs, i.e., $m=2$. It is known that the pair is jointly measurable (i.e., 1-simulable) if and only if $\eta \leq \eta^* = \frac{1}{2}(1+\frac{1}{1+\sqrt{d}})$ \cite{ULMH16,DSFB19}. Hence it follows that for a noise parameter satisfying $1-\frac{1}{2}\frac{d-n}{d-1} \geq \eta > \eta^*$ (which is possible for all $d>2$) we get a set of measurements that is incompatible but nevertheless $n$-simulable (for some $n>1$). Alternatively, consider the full set of $m=4$ MUBs in $d=3$. This set is incompatible if and only if $\eta>(1+3\sqrt5)\approx0.4818$ \cite{DSFB19} while it is $2$-simulable for $\eta \leq \frac{5}{8}=0.625$.

Finally, note that the above construction for $n$-simulability uses a heuristic choice of bases for the compression instrument. We verify below that this choice is sub-optimal.

\paragraph{SDP method.---}
In order to explore more general schemes, involving arbitrary bases for the compression, we now present a numerical approach based on semi-definite programming (SDP). For the example of MUBs, this shows that better schemes are indeed possible.  

 Consider a set of measurements $M_{a|x}$, to which we add noise. For clarity we focus here on the case of white noise, but the technique applies in general. Formally, we consider sets of POVMs of the form: $$M_{a|x}^\eta=\eta M_{a|x}+(1-\eta)\Tr[M_{a|x}]\frac{\id_d}{d} \,.$$
Our goal now is derive a lower bound on $\eta$ for which the set becomes $n$-simulable. In particular, if the lower bound is found to be trivial, i.e., $\eta\geq 1$, we conclude that the original measurements $M_{a|x}$ are $n$-simulable.

We first choose a compression map, consisting of a set of $|\mu|$ instruments $\{ \cE_{\lambda| \mu} \}$. We can now find the maximal value of $\eta$ (i.e., minimise the noise) while ensuring $n$-simulability, optimising over Bob's final measurements, via the following SDP:   
\begin{align}\label{eq: SDP}
    \max_{\eta,\{\tilde{N}_{a|x,\lambda}\},\alpha_\lambda} \;\; & \eta \\
    s.t. \; &\sum_{\lambda=0}^{{d \choose n}-1}\sum_{\mu=0}^{|\mu|-1} \cE_{\lambda| \mu} \tilde{N}_{a|x,\lambda}\cE_{\lambda| \mu}^\dagger  =  M_{a|x}^\eta \nonumber \qquad \forall a,x\\
    &\sum_a \tilde{N}_{a|x,\lambda,\mu} = \alpha_{\lambda,\mu} \id_n \;\;\; \forall x,\lambda,\mu \nonumber \\
    &\tilde{N}_{a|x,\lambda,\mu} \geq 0,\; {\tilde{N}_{a|x,\lambda,\mu}}^\dagger = \tilde{N}_{a|x,\lambda,\mu} \;\;\; \forall a,x,\lambda,\mu \nonumber
\end{align}
To illustrate the relevance of this method, let us consider again the case of a pair of MUBs in dimension $d=3$, i.e., setting $M_{a|x}= \ketbra{\varphi_a^x}{\varphi_a^x}$, and compressing to $n=2$. Here we choose a simple compression channel: we first choose a basis (via a unitary $U_\mu$) and perform a projection onto the $d \choose n$ $n$-dimensional subspaces, denoted by projectors $\Pi_\lambda$. Hence we set $\cE_{\lambda| \mu}= U_\mu \Pi_\lambda$. Optimising over choices of $|\mu|=2$ and $|\mu|=3$ basis, we find noise thresholds of  $\eta\approx0.7803$ and $\eta\approx0.8281$, respectively, hence clearly improving upon the bound of $\eta=5/8$ we got in the analytical construction, see Eq.~\eqref{eqn:claim1}. Moreover, when allowing for more basis (up to $|\mu|= 5$), we could find no improvement.\\

\paragraph{The case of all projective measurements.---} 
So far we analysed sets $\mathcal{M}$ with finitely many measurements. Now we turn our attention to continuous sets of measurements. Precisely, we consider assemblages $\mathcal{M}_{PVM}^\eta$ made of all rank-1 projective measurements subjected to white noise. Note that this automatically extends to all projective measurements which can be obtained from rank-1 projective measurements by post-processing.
Assuming an isotropic noise,
our set is made of all POVMs 
\be\label{eq: all measurements}
 \mathcal{M}_{PVM}^\eta= \{ M_{a|U}^\eta\}_U \quad \text{with} \quad M_{a|U}^\eta = U^\dag M_a^\eta U,
\ee
where $U$ runs through all unitary operators on $\mathds{C}^d$ and $M_a^{\eta}$ is the noisy (or lossy) measurement in the computational basis. On the one hand, the continuous case may seem more complicated as the infinite number of possible measurements cannot be tackled with the SDP of Eq.~\eqref{eq: SDP}. On the other hand, the symmetry of the set helps simplifying the optimal compression scheme as we shall see now.

The set $\mathcal{M}_{PVM}^\eta$ is a particular case of what we call an $\textit{invariant assemblage}$. This is a set $\mathcal{M}$ such that $M_{a}\in \mathcal{M}$ implies that $U^\dag M_{a} U$ is also in $\mathcal{M}$ for all unitaries $U$.
For such an invariant assemblage the choice of the compression bases does not play any role, which leads to the following observation.
\begin{claim}
An $n$-simulable invariant assemblage $\mathcal{M}$  can be compressed with the continuous instrument $\{\cE_V\}$ with CP map density $\cE_V$, such that $\cE_V(\rho)= K_V \rho K_V^\dag$ with
 \be\label{eq: invariant instrument}
 K_V = \sqrt{\frac{d}{n}} \,  \Pi_n V,
\ee
with $V$ running through all unitary operators on $\mathds{C}^d$ and $\Pi_n$ is a projection onto a fixed $n$-dimensional subspace of $\mathds{C}^d$.  In addition, if there exists a measurement $M_a \in \mathcal{M}$ such that $U^\dagger M_a U $ generates all of $\mathcal{M}$ (i.e. the action is \textit{transitive}), then it suffices to find $N_{a|V}$ satisfying $M_a = \int dV \cE_V^*( N_{a| V} ) $, as then $U^\dagger M_a U = \int dV \cE_V^*( N_{a| VU^\dagger} )$.
\end{claim}
Here and below $\dd V =\dd \mu(V)$ is the Haar measure. If there are multiple orbits, that is, a family of measurements $\mathcal{N}$ such that $U^\dagger \mathcal{N} U$ generates all of $\mathcal{M}$, then it suffices to find a measurement $N_{a|V}$ as above for each measurement in $\mathcal{N}$.
A detailed proof of the claim can be found in the Appendix A and we explain only the intuition here.
First, because of the invariance of $\mathcal{M}$ one can freely apply any (random) unitary $V$ to the state before the compression. Thus if the set is compressible with some instrument $\{\cE_\lambda\}$ it is also compressible with an instrument where a random $V$ is applied before $\{\cE_\lambda\}$. Second, one shows that any such instrument can be obtained by post-processing of $\{\cE_V\}$ defined in the claim. Finally, the relation between the compressed POVMs $N_{a|V}$ and $N_{a|VU^\dagger}$ is a simple consequence of Haar measure invariance.

\paragraph{All noisy projective measurements.---} 

We now consider the set of all projective measurements $\mathcal{M}^{\eta}_\text{PVM}$ in Eq.~\eqref{eq: all measurements} subject to white noise
\be
\label{eq: M_aeta}
    M_{a}^\eta= \eta \ketbra{a}{a} + \frac{1-\eta}{d}\id, 
\ee
where $\ket{a}$ denotes the $d$ vectors of the computational basis. We now want to find the $N_{a|V}$ such that $\int dV \cE_V^*( N_{a| V} )$ equals $M_a^\eta$ for the highest value of $\eta$ possible. In fact, the optimal compressed POVM here is given by 
\be\label{eq: optimal N}
N_{a|V} =  \underset{ \tilde{N}_a: \substack{ \, \text{POVM} \\  \,\,\textrm{on} \, \mathds{C}^n}}{\textrm{argmax}} \sum_{a=1}^d  \bra{a} V^\dag \tilde{N}_{a} V \ket{a}
\ee
with $\tilde N_a$ embedded in $\mathds{C}^d$, which can be seen in two steps, c.f. Appendix B for details. First, this choice does result in an operator of the desired form $M_a'=\eta(x) \ketbra{a}{a} + \frac{1-\eta(x)}{d}\id$ with $\eta(x)=\frac{d\,x-1}{d-1}$ and 
\be\label{eq: x general main}
x = \frac{1}{n}  \int \dd V \!\! \underset{\tilde{N}_a: \substack{ \, \text{POVM} \\  \,\,\textrm{on} \, \mathds{C}^n}}{\textrm{max}} \sum_{a=1}^d  \bra{a} V^\dag \tilde{N}_{a} V \ket{a}.
\ee
Second, by construction with the max inside the integral this yields the highest value of $x$, and thus $\eta$, possible (this is precisely the intuition behind the definition~\eqref{eq: optimal N}). This leads to the following full characterisation of $n$-simulability of noisy PVMs in any finite dimensional system.

\begin{claim}
The set of all noisy (white noise) projective measurements  $\mathcal{M}^\eta_\text{PVM}$
in dimension $d$ is $n$-compatible if and only if
\be
\eta \leq \eta_{d\mapsto n} = \frac{d\,  x-1}{d-1}
\ee
with $x$ given in Eq.~\eqref{eq: x general main} 
where $V$ runs through all unitary operators on $\mathds{C}^d$, $\dd V$ is the Haar measure, $\{\ket{a}\}_{a=1}^d$ is the computational basis of $\mathds{C}^d$, and $\mathds{C}^n$ is any $n$-dimensional subspace of $\mathds{C}^d$.
\end{claim}

 \begin{table}[b]
 \label{table 1}
\centering
\begin{ruledtabular}
\begin{tabular}{|l | c c c c c|} 
$n \setminus d$
& $2$ & $3$ & $4$ & $5$ & $6$ \\ \hline
1 &\textit{0.5}& \textit{0.42}& \textit{0.36}& \textit{0.32}& \textit{0.29} \\
2 & & 0.70 & 0.56& 0.48 & 0.42\\
3 & & & 0.77 & 0.64 & 0.55\\
4 && & & 0.81& 0.70 \\
5 && & &     & 0.84 
\end{tabular}
\end{ruledtabular}
\caption{\label{tab:thresholds} Some values of the white noise threshold $\eta_{d\mapsto n}$ for the $n$-simulability of $\mathcal{M}^{\eta}_\text{PVP}$ -- the set  of all projective measurements in dimension $d$. We computed the values with Wolfram Mathematica using the build-in function {\ttfamily CircularUnitaryMatrixDistribution} to sample from the Haar measure, and {\ttfamily SemidefiniteOptimization} to solve the optimal POVMs $\tilde{N}_a$ in Eq.~\eqref{eq: x general main}. The values in italic give the known white noise threshold for the incompatibility of all PVMs and equal to $\eta_{d\mapsto 1} =  (\sum_{k=1}^d \frac{1}{k}-1)/(d-1)$, cf. \cite{wiseman2007steering,uola2014joint}.}
\end{table}

 The POVM maximisation Eq.~\eqref{eq: x general main} is a simple SDP, so the threshold  $\eta_{d \mapsto n}$ can be computed numerically by sampling from the Haar measure (or integrating numerically) and solving the SDP for each $V$. We report some values in Table~\ref{tab:thresholds}.

An upper bound on the threshold can be obtained by applying the Cauchy-Schwarz inequality to the sum in Eq.~\eqref{eq: x general main}, and leads to  
\be
\eta_{d\mapsto n} \leq \frac{d \sqrt{\frac{n+1}{d+1}}-1}{d-1},
\ee
as we show in Appendix C. 
In Appendix D we also derive a lower bound $\eta_{d\mapsto (d-1)} \geq  \frac{d^2-d(1+ \sum_{k=1}^d\frac{1}{k})+1}{(d-1)^2}$ for the case $n=d-1$.

For $n=1$ our considerations reduce to the joint measurability of noisy projective measurements, where the white noise threshold is known $\eta_{d\mapsto1} = (\sum_{k=1}^d \frac{1}{k}-1)/(d-1)$ ~\cite{wiseman2007steering,uola2014joint}. In this case the subspace $\mathds{C}^n$ contains a single state $\ket{\Psi}=V\ket{1}$.  
Accordingly to Eq.~\eqref{eq: x general main} $\eta_{d\mapsto1}$ can be computed from  $x= \int \dd \Psi \max_a |\braket{a}{\Psi}|^2= \frac{1}{d}\sum_{k=1}^d\frac{1}{k}$, 
where $\dd \Psi$ is the uniform measure over states in $\mathds{C}^d$ (invariant under unitrarie), which is equivalent to the derivation in~\cite{wiseman2007steering}.\\

\paragraph{All lossy projective measurements.---}~ 
We now briefly analyse the set of all  projective measurements $\mathcal{M}^{\eta}_\text{PVM}$ in Eq.~\eqref{eq: all measurements} subject to loss
\begin{equation}
    \label{eq: M_loss}
M_{a}^\eta= \eta \ketbra{a}{a} ; \quad  M_\emptyset^{\eta}= (1-\eta)\id 
\end{equation}
This set describes measurements with a limited efficiency -- the additional element $M_\emptyset^\eta$ corresponds to the "no-click" outcome. The required rank-1 form of the operators $M_a^\eta \propto \prjct{a}$ implies that for $a\neq \emptyset$ the POVM element $\hat N_{a|V}$ can only be nonzero if $\cE^*_V(\hat N_{a|V}) \propto \prjct{a}$. In other words, the vector $\ket{a}$ has to be in the $n$-dimensional subspace selected by the instrument, i.e. 
\be
\ket{a} \in \text{span}\{V\ket{1},\dots, V\ket{n}\}.
\ee
This condition is only fulfilled for a set of $V$ of measure zero,
hence $M'_a \propto \prjct{a} \implies M'_a=0$.  We can thus conclude that 
\begin{claim}For any positive efficiency $\eta>0$ the set of all lossy projective measurements $\mathcal{M}_{PVM}^\eta$ is not $n$-simulable for any $n<d$.
\end{claim}

\paragraph{An equivalent definition.---}

The notion of joint measurability has a direct connection to entanglement-breaking properties of quantum channels, cf. \cite{kiukas2017continuous} and \cite{Kogias15,Moroder16}. As a final point we now discuss how this connection naturally extends to $n$-simulability.

A quantum channel $\Lambda$ is said to be \textit{$n$-partially entanglement breaking} ($n$-PEB) if, for all states $\rho$, the Schmidt number of the state $(\Lambda \otimes \id ) [\rho]$
is less than or equal to $n$ \cite{chruscinski2006partially}. In the finite-dimensional setting, \cite{chruscinski2006partially} this is equivalent to the existence of Kraus operators $\{K_\lambda\}$ of $\Lambda$ each with $\text{rank}(K_\lambda)\leq n$. This notion provides an alternative way to define $n$-simulability:
\begin{claim} \label{def: n-PEB}
A measurement assemblage $M_{a|x}$ with finitely many inputs is $n$-simulable if and only if there exists a quantum channel $\Lambda$ that is $n$-PEB and a measurement assemblage $N_{a|x}$ such that 
\be
    M_{a|x} = \Lambda^* \big ( N_{a|x} \big ).
\ee
\end{claim}

For the if direction one uses the singular value decomposition of the rank-$n$ Kraus operators $\{K_\lambda\}$ and $N_{a|x}$ to define the instrument $\{\cE_\lambda\}$ with $n$-dimensional output and the subsequent measurement $N_{a|x,\lambda}$. For the only if direction one can simply view any compression instrument $\{\cE_\lambda\}$ as a channel $\Lambda$ that outputs a quantum system of dimension $n$ and a classical register encoding $\lambda$. This channel is manifestly $n$-PEB. The details of the proof can be found in the Appendix E. Note that we prove the claim for finite amount of classical communication and conjecture it to be true even when the index $\lambda$ runs over a continuous set.

\paragraph{Related concepts.---} We now compare the idea of $n$-simulability with some previously introduced notions. 

First, in \cite{bluhm2018quantum} a related notion of ``$n$-compressibility" of a set of quantum measurements has been proposed. Similarly to our definition
a set of measurements is said to be $n$-compressible if $M_{a|x} = \sum_{\lambda}\cE_\lambda^*(N_{a|x, \lambda})$\footnote{Notably the authors of \cite{bluhm2018quantum} also consider the possibility to bound the amount of classical communication via the number of different instrument outcomes $\lambda$ (the dimension of the classical register output by the instrument).}, but in sharp contrast the compressed measurement has to take the form
$N_{a|x, \lambda} = \tilde{\cE}^*_{|\lambda}(M_{a|x})$.
Here, $\tilde{\cE}_{|\lambda}$ is a set of CPTP maps decompressing the system back into dimension $d$, and the same measurement $M_{a|x}$ has to be used after the decompression. Clearly, a set of $n$-compressible measurements is n-simulable by construction, but not the other way around. In particular, it is known that single measurements can have a larger compression dimension than one \cite{bluhm2018quantum}.

Other works have investigated measurement (in)compatibility in subspaces \cite{LN21,UKD+21}, while Ref.~\cite{Carmeli16} defined a concept of $n$-compatibility considering a scenario where a set of measurements is performed on $n$ copies of a state. As far as we can say, these concepts are unrelated to $n$-simulability.

Finally, it is also worth pointing the approach of ``projective simulability'' discussed in Refs \cite{OGW17+,GBT+17} (see also \cite{Hirsch17}), where one asks whether a set of POVMs can be simulated from projective measurements (of the same dimension) only.

\paragraph{Conclusion.---} We have introduced the concept of $n$-simulability of a set of measurements, motivated by a scenario of compression of quantum information. When full compression is possible (i.e. $n$=1), our notion corresponds to joint measurability. Hence our approach can be used as a quantification of measurement incompatibility in terms of dimension. We discussed a number of examples, providing analytical constructions as well as numerical methods. 

More generally, the concept of $n$-simulability turns out to be connected to several other relevant notions of quantum information theory. First, as we showed above, $n$-simulability relates to partially entanglement-breaking channels. Second, there is a direct connection between $n$-simulability and the notion of genuine high-dimensional steering \cite{designolle2021genuine}, which will be presented in detail in a companion article \cite{Paper2}. These links generalise the well-known connection between joint measurability, steering and quantum channels \cite{uola14,quintino14,uola15}. They also open new questions, for example, whether the dimensionality of a quantum channel could be tested in a partially device-independent manner. 

Beyond quantum compression, our approach also has implications for device-independent quantum information processing. For example, it is clear that a set of measurements that is $n$-simulable is of limited use for randomness certification, as it can lead to (at most) $2\log(n)$ bits of randomness in a black-box setting.

Finally, an intriguing question is whether our approach could be generalised to the case of  continuous variable measurements. It turns out that, in the infinite-dimensional case, the Kraus operators' ranks and entanglement breaking properties of a channel are not anymore tied together \cite{holevo2005separability,shirokov2011schmidt}, opening different possibilities to extend our concept.

\paragraph{Acknowledgments.---} We thank Denis Rosset, Costantino Budroni, Paul Skrzypczyk, Alex Little, and Michalis Skotiniotis for discussions. We acknowledge financial support from the Swiss National Science Foundation (projects 192244, Ambizione PZ00P2-202179, and NCCR SwissMAP). BDMJ acknowledges support
from UK EPSRC (EP/SO23607/1).

\section{Supplementary material}

\subsection{Appendix A: Proof of claim 2}
\label{app: A claim 2}

Here we prove that for a measurement assemblage satisfying $V^\dagger M_{a|U} V  = M_{a|UV}$, we can take our Kraus operators to be of the form $K_V = \sqrt{\frac{d}{n}} \Pi_n V$, where $\Pi_n$ projects onto the first $n$ basis states, and the normalisation condition reads $\int dV K_V^\dagger K_V = \mathbbm{1}$ with $dV$ the Haar measure. We prove this explicitly for measurements $M_{a|U}$ of the above form for clarity of presentation, but the argument holds for any invariant assemblage $\mathcal{M}$.

Suppose we have
\be
M_{a|U} = \sum_\lambda K_\lambda^\dagger ~  N_{a|U, \lambda} \quad K_\lambda,
\ee
for some Kraus operators $K_\lambda$ of rank at most $n$ and measurements $N_{a|U,\lambda}$. We take $\lambda$ as belonging to a countable outcome space, but our proof would extend straightforwardly to the case of continuous instruments (we comment on some subtleties below).

Each Kraus operator admits a singular value decomposition $K_\lambda = U_\lambda D_\lambda V_\lambda$, but by redefining the measurements as $N_{a | U, \lambda} \mapsto U_\lambda^\dagger N_{a | U, \lambda} U_\lambda $ we can without loss of generality take each Kraus operator to be of the form $K_\lambda = D_\lambda V_\lambda$. We can also write this as $K_\lambda = \widetilde{D}_\lambda \Pi_n V_\lambda$, where $\Pi_n$ is a $n \times d$ matrix projecting onto the first $n$ basis states, and $\widetilde{D}_\lambda=\widetilde{D}_\lambda^\dag$ is an $n\times n$ matrix containing the singular values along the diagonal.

Now for each permutation $\pi$ on $n$ elements let $U_\pi = \sum_i \ketbra{\pi(i)}{i}$ be the $n\times n$ unitary matrix permuting the computational basis states. We can thus further write each Kraus operator as follows 
\begin{align}
K_\lambda &= \widetilde{D}_\lambda U_\pi^\dagger U_\pi  \Pi_n V_\lambda \\
&= \widetilde{D}_\lambda U_\pi^\dagger  \Pi_n  \overline{U}_\pi V_\lambda, \label{eq:kraus-op-perm}
\end{align}
where $\overline{U}_\pi= U_\pi\oplus \mathbbm{1}_{d-n}$ is the $d \times d$ unitary matrix with $U_\pi$ acting on the first $n$ basis vectors and the identity on the rest.

In addition, by exploiting the property $V^\dagger M_{a|U} V = M_{a|UV}$ for all $d\times d$ unitaries $V$ we can write 
\be
M_{a|U} =  \int dV \sum_\lambda V^\dagger K_\lambda^\dagger \quad  N_{a|UV^\dagger, \lambda} \quad K_\lambda V,
\ee
where $dV = d\mu(V)$ is the Haar measure. Through the Fubini-Tonelli theorem, we can exchange the sum and integral (we make the natural assumption that $\lambda$ belongs to a $\sigma$-finite measure space).
Then by inserting Eq.~\eqref{eq:kraus-op-perm} and changing variables $ \overline{U}_\pi V_\lambda V \mapsto V$ (using Haar invariance), we obtain the following expression for $M_{a|U}$:
\be
    \sum_\lambda \int dV \quad  V^\dagger \Pi_n^\dagger U_\pi \widetilde{D}_\lambda^\dagger   \quad N_{a|UV^\dagger \overline{U}_\pi  V_\lambda, \lambda} \quad  \widetilde{D}_\lambda U_\pi^\dagger  \Pi_n V.
\ee

We can now average over all permutations $\pi$ to obtain
\begin{align}
    M_{a|U} =   \int dV ~  V^\dagger \Pi_n^\dagger & \bigg [ \frac{1}{n!}\sum_{\lambda, \pi} U_\pi \widetilde{D}_\lambda^\dagger \nonumber\\ \times & N_{a|UV^\dagger \overline{U}_\pi V_\lambda, \lambda} \quad \widetilde{D}_\lambda U_\pi^\dagger \bigg ]  \Pi_n V.
\end{align}
Finally, we can take $K_V := \sqrt{\frac{d}{n}} \Pi_n V$ as Kraus operators satisfying
\be
\int dV K_V^\dagger K_V = \frac{d}{n} \int dV~ V^\dagger \Pi_n V = \mathbbm{1},
\ee
and define 
\be
N_{a|U,V}:= \frac{n}{d\times n!}\sum_{\lambda, \pi} U_\pi \widetilde{D}_\lambda^\dagger \quad  N_{a|UV^\dagger \overline{U}_\pi V_\lambda, \lambda} \quad \widetilde{D}_\lambda U_\pi^\dagger.
\ee
 This defines a valid POVM by virtue of 
\be
\frac{1}{ n!}\sum_{\lambda, \pi} U_\pi \widetilde{D}_\lambda^\dagger  \widetilde{D}_\lambda U_\pi^\dagger = \frac{d}{n}\mathbbm{1},
\ee
where the identity is on $\mathbbm{C}^n$, and we have used the fact that $\widetilde{D}_\lambda$ is diagonal and $\text{Tr}(\sum_\lambda \widetilde{D}_\lambda^\dagger \widetilde{D}_\lambda) = d$.

We have shown that for arbitrary Kraus operators $K_\lambda$ that demonstrate the $n$-simulability of $M_{a|U}$, there always exist measurements $N_{a|U,V}$ such that $K_V$ as defined above also demonstrates the $n$-simulability. Crucially, each $K_V$ has rank $n$, the same as the original Kraus operators $K_\lambda$. This completes the proof. \qed

\subsection{Appendix B: Proof of claim 3}
\label{app: B claim 3}

In this section we show that the measurement $N_{a|V}$ defined in Eq.~\eqref{eq: optimal N} in the main text is optimal for the compression of all noisy projective measurements. The claim 3 then follows straightforwardly.

By claim 2 we know that without loss of generality one can consider compression by means of the continuous instrument $\cE_V$ with Kraus operator density
\be
K_V = \sqrt{\frac{d}{n}}\Pi_n V.
\ee
Furthermore, it also tells us that it is enough to study $N_{a|V}$ achieving the maximal $\eta$ for which the POVM $M_a^\eta$ can be compressed, as it would imply the compression of all measurements $M_{a|U}^\eta$ by symmetry. Hence, we now can focus on finding $N_{a|V}$ such that 
\be
M'_a = \int \dd V \cE_V^*(N_{a|V})
\ee
equals $M_{a}^\eta$ for the largest value $\eta$ possible, i.e. for the lowest amount of white noise.
Precisely, we now show that the measurement
\be\label{eq app: Nav opt}
N_{a|V} =  \underset{ \tilde{N}_a: \substack{ \, \text{POVM} \\  \,\,\textrm{on} \, \mathds{C}^n}}{\textrm{argmax}} \sum_{a=1}^d  \bra{a} V^\dag \tilde{N}_{a} V \ket{a},
\ee
defined in the main text is optimal for the purpose. An attentive reader notes that the right hand side of this equation is not necessary always well defined. For some choices of $V$  there might be several POVMs $\tilde N_a$ that attain the maximum. In these case, we tell the argmax to select such a $\tilde N_a$ at random, defining a valid average POVM. This solves possible issues with argmax and does not affect the following arguments.

The main challenge in proving the optimality of $N_{a|V}$ in Eq.~\eqref{eq app: Nav opt} is to show that it  leads to $M_a'= \eta \ketbra{a}{a} + (1-\eta)\frac{\id}{d}$ of the desired form. 

To prove this, for any fixed unitary $V$ let us also define 
\be
V_{\bm \theta} =  V e^{\ii\, \text{Diag}(\bm \theta)}
\ee
with $\bm \theta = (\theta_1, \dots, \theta_d)$,
and 
\be
V_{\pi} = V U_{\pi}
\ee
with $U_\pi \ket{a} =\ket{\pi(a)}$ permuting the elements of the computational basis. Both $V_{\bm \theta}$ and $V_{\pi}$ are also unitary operators. Furthermore, they satisfy the following properties. For $V_{\bm \theta}$ we have
\be\begin{split}
N_{a|V_{\bm \theta}} &= 
     \underset{ \tilde{N}_a: \substack{ \, \text{POVM} \\  \,\,\textrm{on} \, \mathds{C}^n}}{\textrm{argmax}} \sum_{a=1}^d  \bra{a} V_{\bm \theta}^\dag \tilde{N}_{a} V_{\bm \theta} \ket{a} \\
     &=  \underset{ \tilde{N}_a: \substack{ \, \text{POVM} \\  \,\,\textrm{on} \, \mathds{C}^n}}{\textrm{argmax}} \sum_{a=1}^d  \bra{a} V^\dag \tilde{N}_{a} V \ket{a} \\
     & = N_{a|V}
\end{split}
\ee
since $V_{\bm \theta}\ket{a} = e^{\ii \theta_a} V \ket{a}$  . While for $V_{\pi}$ one finds 
\be\begin{split}
N_{a|V_{\pi}} &= 
     \underset{ \tilde{N}_a: \substack{ \, \text{POVM} \\  \,\,\textrm{on} \, \mathds{C}^n}}{\textrm{argmax}} \sum_{a=1}^d  \bra{a} V_{\pi}^\dag \tilde{N}_{a} V_{\pi} \ket{a} \\
     &=  \underset{ \tilde{N}_a: \substack{ \, \text{POVM} \\  \,\,\textrm{on} \, \mathds{C}^n}}{\textrm{argmax}} \sum_{a=1}^d  \bra{\pi(a)} V^\dag \tilde{N}_{a} V \ket{\pi(a)} \\
     &=  \underset{ \tilde{N}_a: \substack{ \, \text{POVM} \\  \,\,\textrm{on} \, \mathds{C}^n}}{\textrm{argmax}} \sum_{a=1}^d  \bra{a} V^\dag \tilde{N}_{\pi^*(a)} V \ket{a},
\end{split}
\ee
where $\pi^*$ is the inverse permutation of $\pi$. But if $\tilde N_a$ maximizes $\sum_{a=1}^d  \bra{a} V^\dag \tilde{N}_{a} V \ket{a}$ then $\tilde N'_a = N_{\pi(a)}$ maximizes $\sum_{a=1}^d  \bra{a} V^\dag \tilde{N}'_{\pi^*(a)} V \ket{a}$, therefore
\be
 N_{a|V_{\pi}} = N_{\pi(a)|V}.
 \ee
Now we will use the properties $N_{a|V_{\bm \theta}} = N_{a|V}$ and $N_{a|V_{\pi}} = N_{\pi(a)|V}$ to display the symmetries of the operators $M'_a$.  

First by taking an integral with uniform measure $\dd \bm \theta$ for the phases $e^{\ii\, \text{Diag}(\bm \theta)}$ we obtain
\be\begin{split}
&\int \dd \bm \theta \, \cE_{V_{\bm \theta}}^*(N_{a|V_{\bm \theta}}) = 
\int \dd \bm \theta \, \cE_{V_{\bm \theta}}^*(N_{a|V}) \\
& =\int \dd\bm \theta  K_{V_{\bm \theta}}^\dag N_{a|V} K_{V_{\bm \theta}}\\
&= \int \dd \bm \theta e^{-\ii\, \text{Diag}(\bm \theta)}
K_{V}^\dag N_{a|V} K_{V}  e^{\ii\, \text{Diag}(\bm \theta)} \\
& = \int \dd \bm \theta e^{-\ii\, \text{Diag}(\bm \theta)}
\cE_{V}^*(N_{a|V})  e^{\ii\, \text{Diag}(\bm \theta)} \\
& =\sum_a \ketbra{a} \cE_{V}^*(N_{a|V}) \ketbra{a}\\
& = \mathcal{T}_{\{\ket{a}\}} [\cE_{V}^*(N_{a|V})],
\end{split}
\ee
 where  $\mathcal{T}_{\{\ket{a}\}}$ is the twirling map in the computational basis defined in the main text, and we used $\int \dd \theta e^{\ii \theta (n-m)}=\delta_{n,m}$ for integer $n$ and $m$. 
By invariance of the Haar measure and linearity of the maps we then obtain 
\be\begin{split}
M_a' &= \int \dd V \cE_{V}^*(N_{a|V}) \\
&= \int \dd V \cE_{V_{\bm \theta}}^*(N_{a|V_{\bm \theta}}) \\
& = \int  \dd \bm \theta  \dd V \cE_{V_{\bm \theta}}^*(N_{a|V_{\bm \theta}})\\
& = \int  \dd V \mathcal{T}_{\{\ket{a}\}} [\cE_{V}^*(N_{a|V})] \\
&= \mathcal{T}_{\{\ket{a}\}}\left[ \int  \dd V  \cE_{V}^*(N_{a|V}) \right] \\
 & = \mathcal{T}_{\{\ket{a}\}}[M_a'],
\end{split}
\ee
showing that $M_a'$ is diagonal in the computational basis and can be written in the form $M_a'=\sum_{a} x_a \prjct{a}$.
 
Next, for a fixed $a=1,\dots, d$ consider the group of all permutations $\pi \in \Pi_a$ that leave the element $a$ untouched $\pi(a)=a$, it is a subgroup of the permutation group  of size $|\Pi_a|$. With the same idea as above we now compute
\be\begin{split}
\frac{1}{|\Pi_a|} \sum_{\pi \in \Pi_a} \cE_{V_{\pi}}^*(N_{a|V_{\pi}}) &= \frac{1}{|\Pi_a|} \sum_{\pi \in \Pi_a} \cE_{V_{\pi}}^*(N_{\pi(a)|V})\\
&= \frac{1}{|\Pi_a|} \sum_{\pi \in \Pi_a} \cE_{V_{\pi}}^*(N_{a|V}) \\
& = \frac{1}{|\Pi_a|} \sum_{\pi \in \Pi_a} U_\pi^\dag \,  \cE_{V}^*(N_{a|V}) \, U_\pi.
\end{split}
\ee
For the final POVM element this implies
\be\begin{split}
M_a' &= \int \dd V \cE_{V}^*(N_{a|V}) \\ 
& = \int \dd V \cE_{V_\pi}^*(N_{a|V_\pi}) 
\\
 & = \frac{1}{|\Pi_a|} \sum_{\pi\in\Pi_a} \int \dd V \cE_{V_\pi}^*(N_{a|V_\pi}) \\
& = \frac{1}{|\Pi_a|} \sum_{\pi\in\Pi_a} U_\pi^\dag \left(\int \dd V \cE_{V}^*(N_{a|V})  \right) U_\pi \\
& = \frac{1}{|\Pi_a|} \sum_{\pi\in\Pi_a} U_\pi^\dag\,  M'_a \, U_\pi,
\end{split}
\ee
Since $M_a'$ is diagonal, we can now conclude that it is of the form 
\be
M_a' = x_a\prjct{a} + y_a \sum_{a'\neq a} \prjct{a'}
\ee
for all $a$. Furthermore, for any $a'\neq a$ and a permutation $\pi'(a) = a'$
\be\begin{split}
M_{a'}'&=  \int \dd V \cE_{V}^*(N_{a'|V}) \\
&= \int \dd V \cE_{V}^*(N_{\pi'(a)|V})   \\
&= \int \dd V \cE_{V}^*(N_{a|V_\pi})  \\ 
&= \int \dd V U_\pi U_\pi^\dag \cE_{V}^*(N_{a|V_\pi}) U_\pi U_\pi^\dag  \\
& = U_\pi \left( \int \dd V_{\pi} \cE_{V_\pi}^*(N_{a|V_\pi}) \right) U_\pi^\dag \\
& =  U_\pi M'_a U_\pi^\dag.
\end{split}
\ee
Hence, $x_a=x$ and $y_a=y$ are the same for all $a$. Finally, the normalisation $\sum_{a=1}^d M_a'=\id$ implies 
$ y = \frac{1-x}{d-1}$, so we obtain
\be
    M_{a}' = M_a^{\eta} \qquad \text{with} \qquad \eta = \frac{d x - 1}{d-1}.
\ee
Notably, this form of $M_a'$ is true for any choice of POVM $N_{a|V}$ such that $N_{a|V_{\bm \theta}}=N_{a|V}$ and $N_{a|V_{\pi}}=N_{\pi(a)|V}$.

It remains to show that the choice $N_{a|V}$ implies the maximal possible value of $\eta$, or $x$. To show it consider any other choice of measurement $\hat N_{a|V}$ resulting in $\hat M_a'$ of the desired form. For this choice one has
\be\begin{split}
\hat x &= \tr \prjct{a} \hat{M}_a' = \frac{1}{d} \sum_{a=1}^d \tr  \prjct{a} \hat{M}_a' 
\\&= \frac{1}{d} \sum_{a=1}^d \tr\left( \prjct{a}\int \dd V \cE_V^*( \hat N_{a|V} ) \right)\\
  & =  \frac{1}{n}  \int \dd V \sum_{a=1}^d \bra{a}V^\dag \hat N_{a|V} V \ket{a} \\
  & \leq x,
\end{split}
\ee
where
\be\label{eq: x general}
x = \frac{1}{n}  \int \dd V \!\! \underset{\substack{\{N_a\}_{a=1}^d \\ \text{POVM on} \, \mathds{C}_n}}{\textrm{max}} \sum_{a=1}^d  \bra{a} V^\dag N_{a} V \ket{a}
\ee
is precisely the value attained by $N_{a|V}$, which maximises the expression inside the integral by construction. This completes the proof.

Finally, it is worth mentioning that it some cases  it could be interesting to reduce the minimisation over all POVMs $\{\hat{N}_a\}_{a=1}^d$ in the definition of $N_{a|V}$, to only run through projective measurements. Projective measurement can be parametrised by a choice of basis $\{N_b = \prjct{\psi_b}\}_{b=1}^n$ followed by a relabeling $a= A_{b,V}$. The restriction to projective $N_{a|V}$ is in general sub-optimal as supported by numerical results. Nevertheless it defines a valid ansatz
\be\label{eq: projective construction}\begin{split}
\{ N_{b|V}'\}_{b=1}^n &=  \underset{\substack{ \{\ket{\psi_b} \}_{b=1}^n \\ \text{basis on} \, \cH_n}}{\textrm{argmax}} \sum_{b=1}^n \max_{a\in\{1,\dots,d\} }| \bra{a} V^\dag  \ket{\psi_b}|^2\\
A_{b,V}&= \underset{a \in\{1,\dots,d\}}{\textrm{argmax}} \bra{a} V^\dag N_{b|V}' V \ket{a},
\end{split}
\ee
leading to the POVM $\{\hat N_{a|V}' \}_{a=1}^d$ with
\be
 \hat N_{a|V}' = \sum_{b=1}^n \delta_{a,A_{b,V}} N'_{b|V}.
\ee
By repeating the above arguments, one sees that this construction also results in $M_a'= x'\prjct{a}+ y'\sum_{a'\neq a} \prjct{a}$ with 
\be\label{eq: x projective}
x' = \frac{1}{n}  \int \dd V \!\! \underset{\substack{\{\ket{\psi_b}\}_{b=1}^n \\ \text{basis on} \, \cH_n}}{\textrm{max}} \sum_{b=1}^n \max_{a\in\{1,\dots,d\}}  |\bra{a} V^\dag \ket{\psi_b}|^2 \leq x.
\ee

\subsection{Appendix C: Upper bound on n-simulability threshold of all noisy PVMs}
\label{app: C upper bound}

In this section we derive a upper bound on the white noise $n$-simulability threshold of all projective measurements $\eta_{d\mapsto n}$. The starting point is to using the Cauchy-Schwarz inequality to upper bound $x$ in Eq.~(13) of the main text. Precisely, for two vectors of operators $\bm A = (\dots,A_i,\dots)$ and $\bm B$ one may defines an inner product $(\bm A, \bm B) = \sum_i \tr A_i  B_i^\dag$. Then Cauchy-Schwarz inequality then implies $(\bm A, \bm B) \leq \sqrt{\sum_i \tr A_i A_i^\dag  \sum_j\tr B_j B_j^\dag}$. In our case, 
\be\nonumber\begin{split}
\sum_{a=1}^d \bra{a} V^\dag N_a V \ket{a} 
&= \sum_{a=1}^d \tr \left ( \Pi_n V \ketbra{a}{a} V^\dag \Pi_n ~ N_a  \right ) \\
&\leq \sqrt{\sum_{a=1}^d \tr (\Pi_n V \ketbra{a}{a} V^\dag \Pi_n)^2 \sum_{a=1}^d \tr N_a^2} \\
 &\leq  \sqrt{\sum_{a=1}^d |\bra{a}V^\dag \Pi_n V \ket{a}|^2 \sum_{a=1}^d \tr N_a}\\
& = \sqrt{   \sum_{a=1}^d |\bra{a}V^\dag \Pi_n V \ket{a}|^2 \, n},
\end{split}
\ee
where we used $N_a^2 \leq  N_a$, enforced by $0\leq N_a \leq \id$, and $\sum_a N_a = \id_n\oplus 0_{d-n}$.
Plugging in the definition of  $x$ one obtains
\begin{align}
\nonumber x &= \frac{1}{n}  \int \dd V \!\! \underset{\tilde{N}_a: \substack{ \, \text{POVM} \\  \,\,\textrm{on} \, \mathds{C}^n}}{\textrm{max}} \sum_{a=1}^d  \bra{a} V^\dag N_{a} V \ket{a}\\
&\label{eq: ub x 1}\leq  \sqrt{\frac{1}{n}}  \int \dd V  \sqrt{ \sum_{a=1}^d |\bra{a}V^\dag \Pi_n V \ket{a}|^2}\\
&\label{eq: ub x 2} \leq \sqrt{ \frac{1}{n}\int \dd V \sum_{a=1}^d |\bra{a}V^\dag \Pi_n V \ket{a}|^2}, \\
&= \sqrt{ \frac{1}{n} \sum_{a=1}^d\int \dd V  |\bra{a}V^\dag \Pi_n V \ket{a}|^2}, \\
&=\sqrt{ \frac{d}{n}\int \dd \Psi |\bra{\Psi} \Pi_n \ket{\Psi}|^2}, \label{eq: ub nice form}
\end{align}
where the last inequality follows by Jensen's inequality and the concavity of the square root function. Notably both upper bounds (\ref{eq: ub x 1})  and (\ref{eq: ub nice form}) do not contain an optimisation.  

To compute the last integral let us define
\begin{align}
 \Theta (d,n) &= \int \dd \Psi \abs{\bra{\Psi}\Pi_n\ket{\Psi}}^2 \\
  &= \text{Tr}\bigg ( \int \dd\Psi ~ \ketbra{\Psi} \otimes \ketbra{\Psi} ~ \Pi_n \otimes \Pi_n \bigg ).
\end{align}
Now recall that $\int \dd\Psi ~ \ketbra{\Psi} \otimes \ketbra{\Psi}$ denotes the normalised projector onto the symmetric subspace $\frac{2}{d(d+1)}\Pi_\text{sym} = \frac{(\mathbbm{1}+S)}{d(d+1)}$, with $S$ the swap operator \cite{harrow2013church}. Hence we can write
\begin{align}
 \Theta (d,n) &=\text{Tr}\bigg ( \frac{(\mathbbm{1}+S)}{d(d+1)} ~ \Pi_n \otimes \Pi_n \bigg ) \\
 &=\frac{1}{d(d+1)}\bigg [ \text{Tr} ( \Pi_n )^2 + \text{Tr} ( \Pi_n^2 ) \bigg ] \\
 &= \frac{n(n+1)}{d(d+1)},
\end{align}
where in the second line we used the fact that $\text{Tr} ( A \otimes B S) = \text{Tr}(AB)$.

By virtue of Eq.~\eqref{eq: ub nice form} we thus find $ x \leq \sqrt{\frac{n+1}{d+1}}$, and
\be
\eta_{d\mapsto n} \leq \frac{d \sqrt{\frac{n+1}{d+1}}-1}{d-1}.
\ee

\subsection{Appendix D: Compression from $d$ to $n=d-1$, a lower bound}

We here focus on the $d-1$-simulability of all noisy projective measurement. As shown in the main text, finding the white noise threshold $\eta$ boils down to solving the integral of Eq.~(13). In particular one has to solve the maximisation 
\be\label{eq: maxim last}
X(V)=\underset{ \tilde{N}_a: \substack{ \, \text{POVM} \\  \,\,\textrm{on} \, \mathds{C}^{d-1}}}{\textrm{max}} \sum_{a=1}^d  \bra{a} V^\dag \tilde{N}_{a} V \ket{a}.
\ee

Here we change the notation slightly by considering the subspace $\mathds{C}^{d-1}=\text{span}\{\ket{2},\dots {\ket{d}} \}$, so that the POVM elements sum up to  $\sum_{a=1}^d \tilde{N} = \Pi_{d-1}$ identity on this subspace and fulfill $\tilde{N}_a\ket{1}=0$.

It is now convenient to use the composite parametrization of the unitary group introduced in \cite{spengler2012composite}
\be\label{eq: Spengler param}
V^\dag =\left[\prod_{m=1}^{d-1}\left(\prod_{k=m+1}^d e^{\ii \lambda_{k,m} P_k} e^{\ii \lambda_{m,k} Y_{m,k}}\right) \right]\cdot \left[\prod_{\ell=1}^d e^{\ii \lambda_{\ell, \ell} P_\ell}\right]
\ee
with the hermitian opertators
\be
Y_{m,k} = \ii (\ketbra{k}{m}-\ketbra{m}{k}) \qquad  P_\ell = \ketbra{\ell},
\ee
$d^2$ real parameters $\lambda_{m,k} \in [0,2\pi]$ for $m\geq k$ and $\lambda_{m,k} \in [0,\frac{\pi}{2}]$ for $m< k$, integer indices $m,k =1,\dots,d$, and all terms appearing in the products from left to right for increasing index. In~\cite{spengler2012composite} the expression of the Haar measure for the above parametrization is also derived, it reads
\be\label{eq: Haar spengler}
\dd V = \frac{1}{N_d} \prod_{m-1}^{d-1}\prod_{k=m+1}^d \sin(\lambda_{m,k}) \cos^{2(k-m)-1}(\lambda_{m,k}) \prod_{\ell,j} \dd \lambda_{\ell,j}
\ee
with the normalization constant $N_d$ that we are not going to need.

This parametrization of Eq.~\eqref{eq: Spengler param} is convenient because it allows us to decompose 
\be
V^\dag =\left(\prod_{k=2}^d e^{\ii \lambda_{k,1} P_k} e^{\ii \lambda_{1,k}  Y_{1,k}}\right) \overline{W}^{(d-1)} e^{\ii \lambda_{1,1} P_1 }
\ee
with 
\be
\overline W^{(d-1)} = \left[\prod_{m=2}^{d-1}\left(\prod_{k=m+1}^d e^{\ii \lambda_{k,m} P_k} e^{\ii \lambda_{m,k} Y_{m,k}}\right) \right]\cdot \left[\prod_{\ell=2}^d e^{\ii \lambda_{\ell, \ell} P_\ell}\right]
\nonumber
\ee
which acts trivially on $\ket{1}$. 

Denoting $U= \left(\prod_{k=2}^d e^{\ii \lambda_{k,1} P_k} e^{\ii \lambda_{1,k}  Y_{1,k}}\right)$ we can now rewrite the maximization Eq~\eqref{eq: maxim last} with the $V$ expressed in terms of the parameters $\lambda_{m,k}$. Noting that $ e^{\ii \lambda_{1,1} P_1 } \tilde{N}_a e^{-\ii \lambda_{1,1} P_1 } =\tilde {N}_a$ we obtain
\be\begin{split}
X(V) &=\underset{ \tilde{N}_a: \substack{ \, \text{POVM} \\  \,\,\textrm{on} \, \mathds{C}^{d-1}}}{\textrm{max}} \sum_{a=1}^d  \bra{a} U \overline{W}^{(d-1)} \tilde{N}_{a} \overline{W}^{(d-1)\dag } U^\dag \ket{a} \\
& = \underset{ \tilde{N}_a: \substack{ \, \text{POVM} \\  \,\,\textrm{on} \, \mathds{C}^{d-1}}}{\textrm{max}} \sum_{a=1}^d  \bra{a} U \tilde{N}_{a} U^\dag \ket{a}
\end{split}
\ee
because the transformation $\tilde{N}_a \to \overline{W}^{(d-1)\dag } \tilde{N}_a \overline{W}^{(d-1)}$ does not affect the optimization.

\paragraph{Lower bound--} Next, to obtain a lower bound on $x$ we relax the maximisation by fixing 
\be
\tilde{N}_a = \begin{cases}
0 & a= 1 \\
\prjct{a} & 2\leq a \leq d 
\end{cases}
\ee
in the last equation. This choice implies 
\be
X(V)\geq X'(V) = \sum_{a=2}^{d} |\bra{a} U \ket{a}|^2.
\ee
Let us now compute
\be
|\bra{a} U \ket{a}| = \left|\bra{a} \left(\prod_{k=2}^d e^{\ii \lambda_{k,1} P_k} e^{\ii \lambda_{1,k}  Y_{1,k}}\right) \ket{a} \right|
\ee
for $a\geq 2$. Using 
\be\begin{split}
\bra{a} \left(\prod_{k=2}^{a-1} e^{\ii \lambda_{k,1} P_k} e^{\ii \lambda_{1,k}  Y_{1,k}}\right) &= \bra{a}\\ 
 \left(\prod_{k=a+1}^{d-1} e^{\ii \lambda_{k,1} P_k} e^{\ii \lambda_{1,k}  Y_{1,k}}\right) \ket{a} &= \ket{a},
\end{split}
\ee
one obtains
\be\begin{split}
|\bra{a} U \ket{a}| &= \left|\bra{a} e^{\ii \lambda_{a,1} P_a} e^{\ii \lambda_{1,a}  Y_{1,a}}\ket{a} \right|\\
&=   \left|\bra{a} e^{\ii \lambda_{1,a}  Y_{1,a}}\ket{a} \right| \\
&= |\cos(\lambda_{1,a} )|,
\end{split}
\ee
and
\be
X'(V)= \sum_{k=2}^d \cos^2(\lambda_{1,k}).
\ee

To express the lower bound 
\be
x = \frac{1}{d-1}\int \dd V X(V) \geq  \frac{1}{d-1}\int \dd V X'(V)
\ee
it thus remains to compute the integrals
\be\begin{split}
I_k &= \int \dd V \cos^2(\lambda_{1,k})\\
& = \frac{\int_{0}^{\pi/2} \sin(\lambda_{1,k}) \cos^{2k-1}(\lambda_{1,k}) \dd \lambda_{1,k} }{\int_0^{\pi/2} \sin(\lambda_{1,k}) \cos^{2k-3}(\lambda_{1,k}) \dd \lambda_{1,k} }\\
&= 1- \frac{1}{k}
\end{split}
\ee
where we used the expression of the Haar measure in Eq.~\eqref{eq: Haar spengler}.
Taking the sum over $k$ one obtains 
\be
x\geq \frac{\sum_{k=2}^d I_k}{d-1} =  \frac{\sum_{k=1}^d I_k}{d-1} = \frac{d-\sum_{k=1}^d\frac{1}{k}}{d-1}.
\ee 
With the claim 3 we finally gets to the lower bound reported in the main text
\be
\eta_{d\mapsto d-1} \geq  \frac{d^2-d(1+ \sum_{k=1}^d\frac{1}{k})+1}{(d-1)^2}.
\ee

\subsection{Appendix E: Proof of claim 5}

\label{app : F claim 5}

Below we prove the equivalence of the claim 5, which proposes that the two following definitions of n-simulability are equivalent for a finite $d$: 
\be\begin{split}
&(1)\qquad M_{a|x}= \sum_{\lambda} \cE_\lambda^*(N_{a|x,\lambda})\\
&(2)\qquad M_{a|x}= \Lambda^*(N'_{a|x})
\end{split}
\ee
In the former case $(1)$, $\{\cE_\lambda\}$ made of CP-maps $\cE_\lambda:B(\mathds{C}^d) \to B(\mathds{C}^n)$ and $N_{a|x,\lambda}$ is a POVM on $B(\mathds{C}^n)$. In the latter (2), $\Lambda$ is a $n$-partially entanglement breaking channel, for which the dimension of the output quantum system is not a priori specified, and $N'_{a|x}$ is a subsequent POVM. As mentioned in the main text the $n$-PEB map $\Lambda$ has Kraus operators or rank-$n$ at most.

We start by showing that $(1) \implies (2)$, to do so let us introduce a classical register $C$ that encodes the classical output of the instrument $\{\cE_\lambda\}$. We can then associate the instrument with a channel
\be
\Lambda: \rho \mapsto \sum_\lambda \cE_\lambda[\rho]_Q \otimes \prjct{\lambda}_C,
\ee
 that outputs a quantum-classical state with a quantum system of dimension $n$ and a classical register $\lambda$. By construction $\Lambda$ is $n$-PEB. In turn, the measurement $N_{a|x,\lambda}$ conditioned on the the classical label $\lambda$, can be viewed as a POVM $N'_{a|x}$ on the quantum-classical system $QC$. Hence, an $n$-simulable assemblage accordingly to Def. (1) is also $n$-simulable accordingly to Def. (2).\\

To show $(2)\implies (1)$ consider a $n$-PEB channel $\Lambda$ and a measurement assemblage $M_{a|x}$. The channel $\Lambda$ admits a Kraus representation \{$K_\lambda \}$ with Kraus operators of rank $n$. Each Kraus operator then admits the singular value decomposition
\be
K_\lambda =U_\lambda D_\lambda V_\lambda^\dag,
\ee
where  $U_\lambda^\dag U_\lambda=\id$ and $V_\lambda V_\lambda^\dag=\id$, and $D_\lambda:\mathds{C}^n \to \mathds{C}^n$
with $\id \geq D_\lambda \geq 0$ is diagonal in the preferred basis. The Kraus operators satisfy $\id =\sum_\lambda K^\dag _\lambda K_\lambda = \sum_\lambda V_\lambda D_\lambda^2 V_\lambda^\dag$.

Let us now define a quantum instrument  $\{\cE_\lambda\}$ as
\be
\cE_\lambda [\rho] = D_\lambda V_\lambda^\dag\,  \rho \, (D_\lambda V_\lambda^\dag )^\dag.
\ee
It is a valid instrument $\sum_\lambda  (D_\lambda V_\lambda^\dag)^\dag (D_\lambda V_\lambda^\dag) =\id$ and manifestly outputs a quantum system of dimension $n$. Furthermore, with the measurement $N'_{a|x}$ let us define 
\be
N_{a|x,\lambda} = U_\lambda^\dag N'_{a|x} U_\lambda,
\ee
which also acts on the $n$-dimensional system. It is straightforward to verify that 
\be\begin{split}
\Lambda^*\Big(N'_{a|x} \Big) &= \sum_\lambda V_\lambda D_\lambda U_\lambda^\dag N'_{a|x} U_\lambda D_\lambda V_\lambda^\dag   \\
&= \sum_\lambda \cE_\lambda^* \Big( N'_{a|x, \lambda} \Big).
\end{split}
\ee
This completes the proof.

\bibliographystyle{unsrt}
\bibliography{References}

\end{document}